\documentclass[doublecol]{epl2} 
\usepackage{epsfig}
\usepackage{pifont}
\usepackage{amssymb}
\usepackage{bm}

\newcommand{\be}{\begin{equation}}
\newcommand{\ee}{\end{equation}}

\newcommand{\degree}{\ensuremath{^\circ}}

\title{Multiple time scale dynamics in the breakdown of superhydrophobicity}
\shorttitle{Dynamics of the spontaneous breakdown of superhydrophobicity}

\author{C. Pirat$^1$, M. Sbragaglia$^1$, A. M. Peters$^2$, B. M. Borkent$^1$, R. G. H. Lammertink$^2$, M. Wessling$^2$ and D. Lohse$^1$ }

\institute{$^{1}$ Physics of Fluids, Department of Applied Physics, University of Twente,\\ P.O. Box 217, 7500 AE Enschede, The Netherlands.\\
$^2$Membrane Technology Group, Department of Chemical Engineering, University of Twente,\\ P.O. Box 217, 7500 AE Enschede, The Netherlands.}


\abstract{
Drops deposited on rough and hydrophobic surfaces can stay suspended with gas pockets underneath the liquid, then showing very low hydrodynamic resistance. When this superhydrophobic state breaks down, the subsequent wetting process can show different dynamical properties. A suitable choice of the geometry can make the wetting front propagate in a stepwise manner leading to {\it square-shaped} wetted area: the front propagation is slow and the patterned surface fills by rows through a {\it zipping} mechanism. The multiple time scale scenario of this wetting process is experimentally characterized and compared to numerical simulations.}

\begin{document}

\maketitle

When a liquid droplet is deposited on hydrophobic micro-structured materials, it can bead off and stay suspended with very high contact angles (Cassie-Baxter state, hereafter CB) and very low hydrodynamic resistance ("Lotus effect") \cite{Degennes03,Barth97,Bico99,Quere}. This property makes such materials useful for a wide series of applications ranging from coating to microfluidics \cite{Squires05,Lauga05,Otten04}. However, when superhydrophobicity breaks down, fluid can enter and fill the micro-structures, resulting into a smaller effective contact angle (Wenzel state, hereafter W). In some situations this transition from the CB to the W state is highly desirable. An example is provided by heterogeneous porous catalysts, where superhydrophobicity is an unwanted effect as it reduces the contact area \cite{Feng02}. Understanding the mechanisms triggering the transition and characterizing its dynamical properties is crucial. The triggering mechanisms for the transitions are still widely debated in the literature \cite{Patankar}. This is due to the presence of energy barriers encountered when passing from CB to W that may strongly depend on the filling procedure. This means that even when the Wenzel state has lower energy (lower contact angle), the transition is not always spontaneous and has to be triggered, for example, by pressing on the drop \cite{Bico99} or by using volumetric forces such as gravity \cite{Patankar}. However, under certain conditions \cite{Sbragaglia07}, local spontaneous infiltrations can be achieved. Whether the liquid is able to then spread and fill the micro-structures depends of the surface patterning. In particular, when the surface roughness becomes comparable to an intrinsic value (critical value), a smooth filling is replaced by a step-like process in which rows of the structured surfaces fill through a zipping mechanism \cite{Sbragaglia07}. This gives rise to slowing down in the system and the emergence of a dynamics with different time scales. The aim of this paper is to characterize this scenario experimentally and with the help of numerical simulations.

\begin{figure*}[t]
\center
\includegraphics[height=7cm,keepaspectratio]{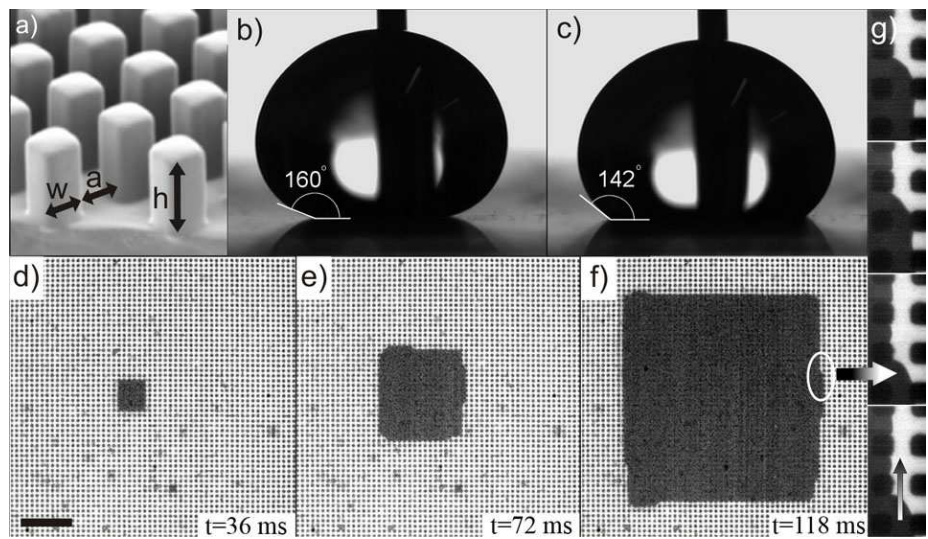}
\caption{\label{fig1}(a) Scanning electron microscopy (SEM) picture of the patterned surface. Here $h=10\,\mu m$, $w=5\,\mu m$, and $a=5\,\mu m$. The water droplet spontaneously transits from the metastable CB (b) to the W state (c), resulting in a lower contact angle (AVI movie, 0.25 Mb). (d-f) The microstructure underneath the liquid fills in a few hundreds of milliseconds through a square-like pattern involving a {\it zipping} mechanism (AVI movie, 4.4 Mb). (g) Enlargement of the zipping mechanism: a row perpendicular to the front direction is rapidly filled. The bar in (d-f) indicates $100\,\mu m$.}
\end{figure*}

\begin{figure*}[t]
\center
\includegraphics[height=15cm,keepaspectratio,angle=270]{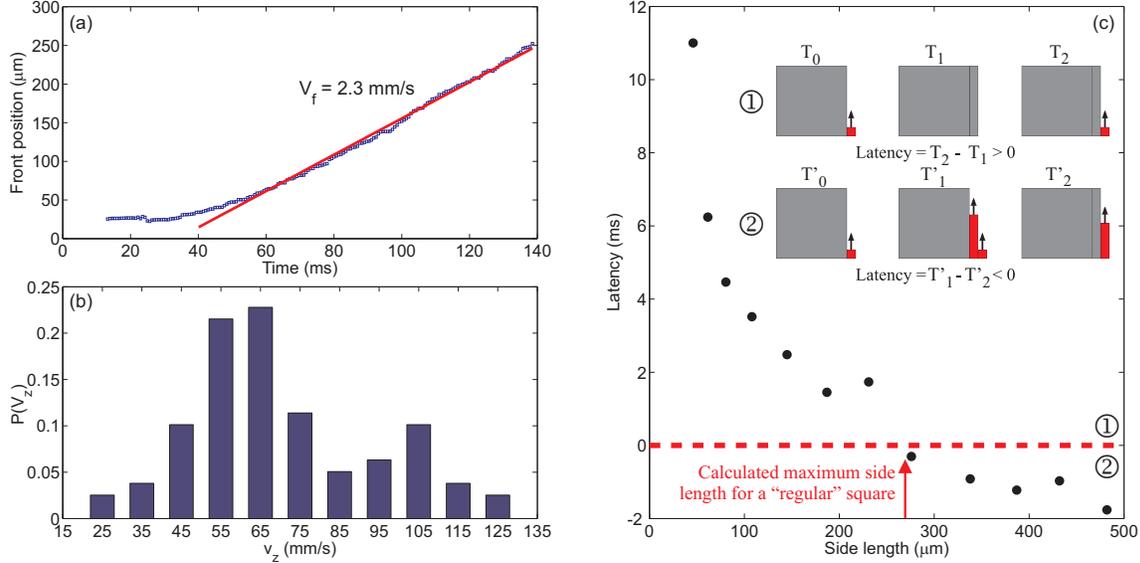}
\caption{\label{zipping10} (a) Temporal evolution of the front position. Values have been averaged over the 4 sides of the square-shaped pattern, starting from its center. (b) Probability distribution of the zipping velocity $V_z$. (c) Measured latency as a function of the side length. How latencies have been calculated is explained in the text. Acording to the theory, the latency becomes negative at a side length of $270\,\mu m$ (see arrow). As observed in the corresponding movie, beyond this value the pattern is not a perfect parallelepiped anymore.}
\end{figure*}

\begin{figure*}[t!]
\center
\includegraphics[width=10cm,keepaspectratio]{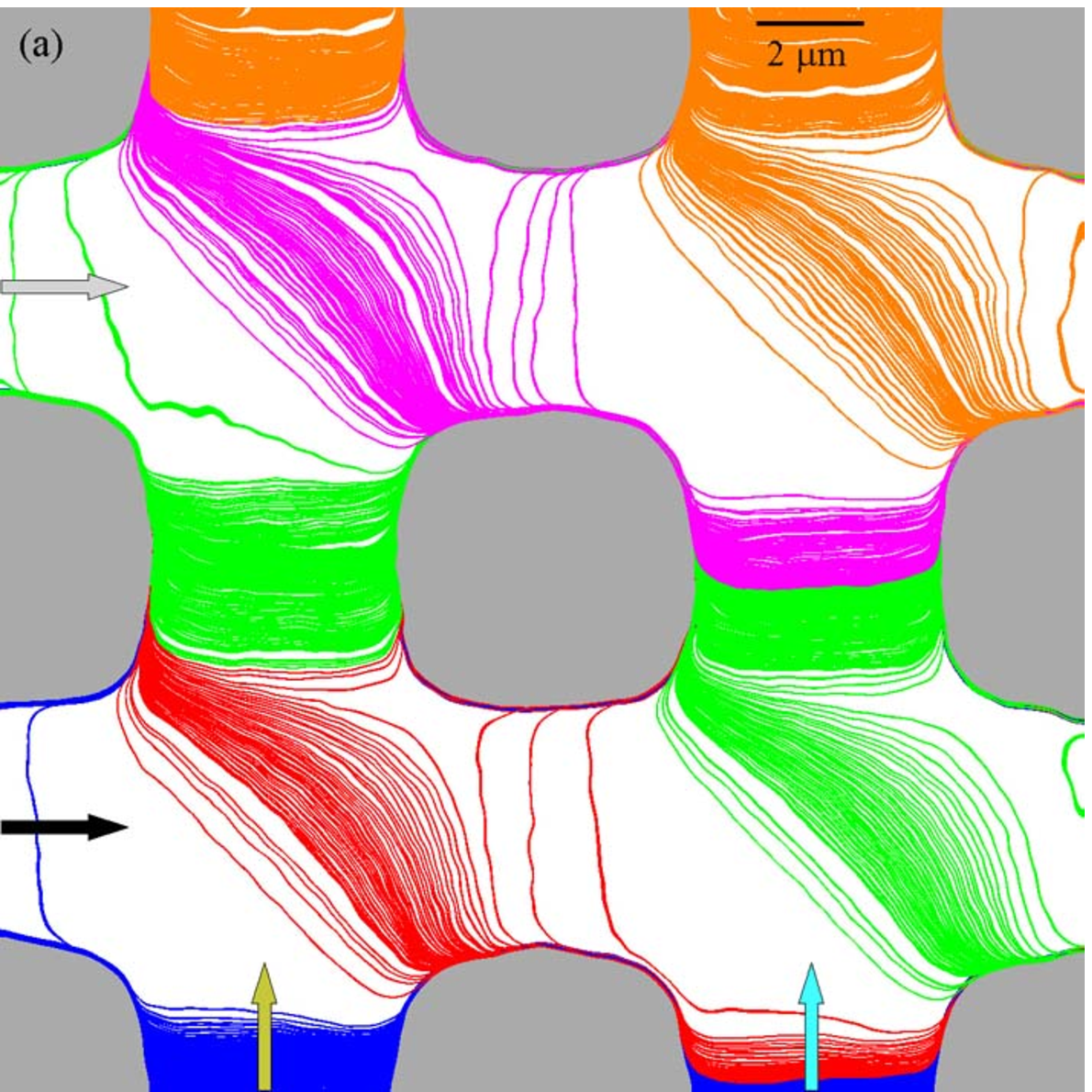}

\includegraphics[width=12.0cm,keepaspectratio]{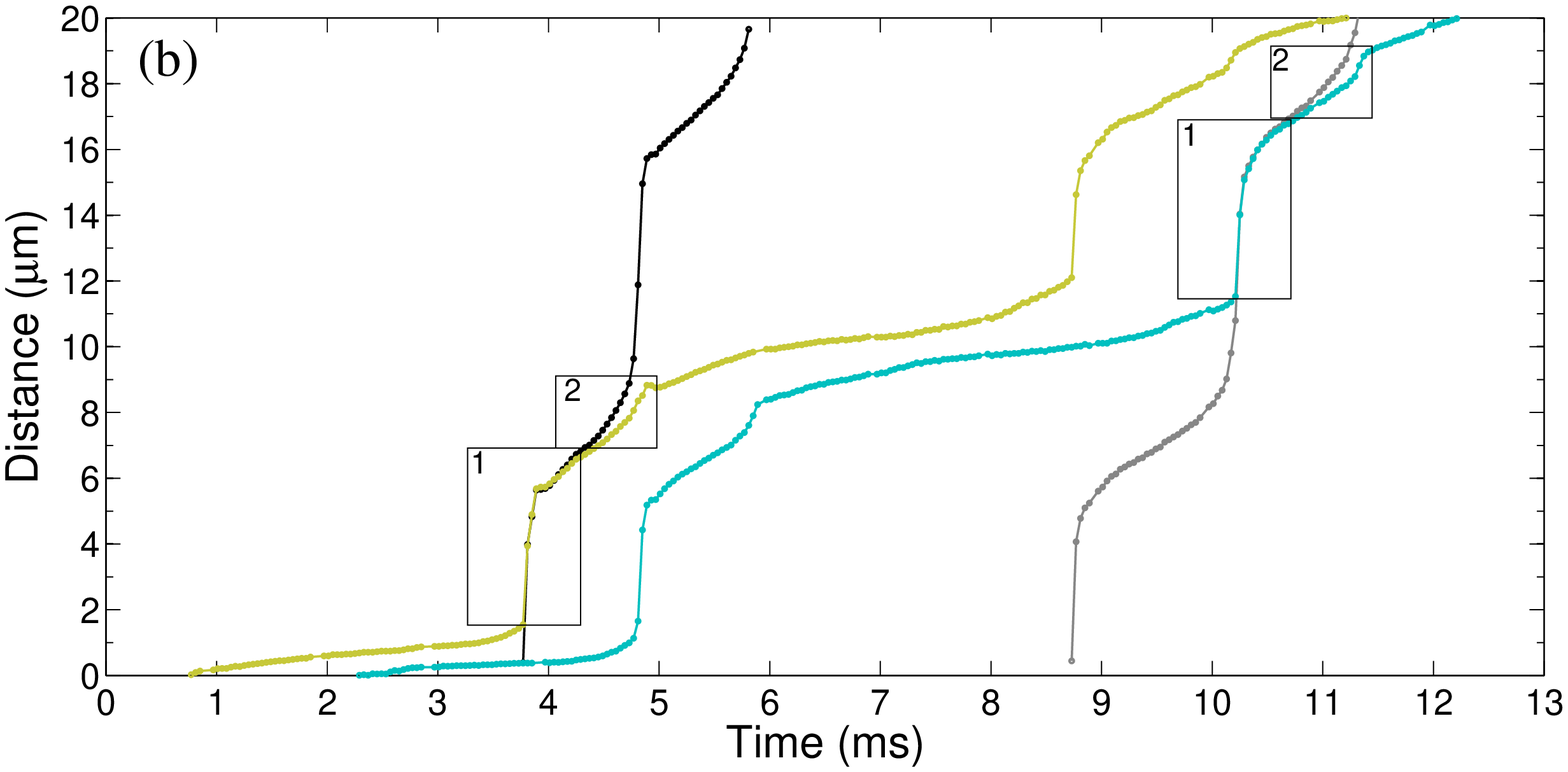}
\caption{\label{zipping100} (a) Details of the zipping process in the experiment. Liquid-gas interface positions obtained every $20\,\mu s$ are simultaneously drawn. The filling begins at bottom-left corner. The main front (responsible for the macroscopic square shape) is from bottom to top. The zipping front (in a row to be filled) is from left to right. Color changes after an interface has detached from a pillar; in chronological order: blue, red, green, pink, orange (AVI movie, 2.6 Mb). (b) Interface position as a function of time along the 4 straight lines suggested by the arrows in (a). Curve and arrow colors match. Labelled rectangles correspond to different steps in the dynamics, as explained in the text.}
\end{figure*}

For the experiments presented here we choose micrometer-sized square pillars set on a regular square lattice with height $h=10\,\mu m$, width $w=5\,\mu m$, spacing $a=5\,\mu m$, and hence $d=w+a=10\,\mu m$ as lattice wavelength (see fig.~\ref{fig1}(a)). Fabrication of highly precise and controllable micro-structured surfaces becomes possible through a micro-molding technique \cite{Jong06}. As material, we choose a polymer fabricated from a solution of styrene-butadiene-styrene (a block-copolymer commercially available as ``kraton'') dissolved in toluene, and then applied on a silicon mold prepared by standard photolythography and etching techniques. After evaporation, the resulting micro-patterned thin polymeric film is bonded to a $170\,\mu m$ thick microscope glass slide. Equilibrium contact angle for water is of $\theta\sim\,160\,\degree$ in the metastable CB state and $\theta\sim\,142\,\degree$ in the W state (fig.~\ref{fig1}(b-c)), whereas contact angle for a smooth surface made of the same material is of $\theta_f\sim\,100\,\degree$.

Water droplets are softly deposited on the surface with a syringe pump (PHD 2000, Harvard Appartus GmbH, March-Hugstetten, Germany) at a very low flow rate ($5\,\mu l/min$) from the outlet of a vertical thin tubing (outer diameter $0.158\,mm$). The outlet is set parallel to the flat film, $2\,mm$ above it. A typical droplet grows slowly and reaches the dry surface within a minute. Then the flow is stopped and a stable drop of contact area $\sim\,1\,mm^2$ in the CB state is observed. Observations are performed from the bottom of the film with an inverted microscope (Axiovert 40 CFL, Carl Zeiss BV, Weesp, The Netherlands). Proper illumination is obtained with a fiber lamp (ILP-1, Olympus, Zoeterwoode, The Netherlands), combined with a high-speed charged coupled device camera (APX-RS, Photron Limited, UK). Focusing in the material features is achieved with a piezoelectric objective-lens positioning system with reproducibility below $100\,nm$ (MIPOS 500, Piezosystem Jena GmbH, Jena, Germany). The translucent film is only $\sim\,40\,\mu m$ thick, allowing for high quality optical imaging of the CB to W transition through its bottom.

The large scale dynamics is captured with a $10\times$ magnification objective at 5 kfps. Smaller details are resolved at 50 kfps with an oil immersion Plan-Apochromat $100\times$ objective, of numerical aperture $NA=1.4$, allowing for a few microns thick image plane. 

Submicron resolution is achieved with the help of a home-made image post-processing technique. First a movie with region of interest $20\,\mu m\times20\,\mu m$ is decomposed in an image sequence. For each image, the initially $117\times117$ pixels window undergoes a $10\times$ oversampling up to $1170\times1170$ pixels. Next, a FFT bandpass filter consisting in a twofold $2d$ Gaussian filtering in the Fourier space is applied. The low-pass component acts as a small structure smoother, whereas the high-pass one corrects large scale illumination inhomogeneities and increases contour contrast. Finally, adaptive intensity threshold is properly applied and interface position detected. Attention should be paid on the fact that this method is sensitive to illumination variations through the image sequence, and may result from time to time in a slight shift of the detected interface position. Moreover, shadowing caused by pillars yields an imprecise detection of the interface position in the vicinity of the pillars, although the main part is accurately detected within $100\,nm$.

After depositing the drop on the surface, a spontaneous breakdown of the metastable CB state starts usually within a minute. From a local infiltration point, a rapid spreading (responsible for the lateral front, hereafter called ``main front'') develops through the square lattice posts. The details of this local infiltration point are sensitively dependent on the height (i.e. roughness): the higher the posts, the larger the time needed to achieve that. This is probably due to the presence of some local energy barrier \cite{Patankar} whose value is increasing with surface roughness. In this way, for large $h$, the infiltration threshold cannot be achieved spontaneously and some local trigger mechanism is therefore necessary. Once the liquid has entered the micro-structures, the dynamics of liquid moving within the pillars can be understood in terms of energy balance between the pulling mechanism of the liquid-gas interface on the top of the pillars (energy gain) and the repulsive wetting of the hydrophobic walls (energy loss). Therefore, as shown in \cite{Sbragaglia07}, whether the front propagates depends on the ratio $a/h$. More precisely, if $\theta<\theta_c$ with
\be \label{thetac}
\cos \theta_{c}=-1+\frac{2}{2+a/h}
\;\;,
\ee
then filling can occur. The contact angle $\theta_f$ for water on flat kraton is $\theta_f\approx100\degree$, thus very close to the critical contact angle $\theta_c=101.5\degree$ set by $a=5\,\mu m$ and $h=10\,\mu m$. Therefore, the dynamics is slow and a square-shaped pattern grows in a stepwise manner, by successively filling the rows along the 4 sides (zipping regime, fig.~\ref{fig1}(d-g)). It has also been pointed out in \cite{Sbragaglia07} that this critical slowing down (involved when $\theta_f$ becomes close to $\theta_c$) results in large dispersion of the measured front velocities, carried out from several experiments with the same sample, although in each case the front velocity remains almost constant during the filling.
It can be noticed that, although averaged front velocities vary from one experiment to the other close to the critical point, the main features persist, provided that experiments are performed close enough to the critical point (i.e. the zipping persists). The question we address in this letter is: what characterizes the dynamics during the invasion process? 

To answer this question, we have first performed large scale measurements. Figure~\ref{fig1}(d-f) shows snapshots during filling while fig.~\ref{zipping10}(a) shows the main front position as a function of time in a typical experiment. When the filled area only extends over few pillars the dynamics undergoes a transitional regime, also observed in the numerical simulations. Afterwards, the front velocity is found to be $V_f=2.3\,mm/s$. Since $V_f$ is almost constant and, at the same time, the square-shaped pattern remains, one may wonder how the zipping process adjusts. Figure~\ref{zipping10}(b) shows the probability distribution of the zipping velocity $V_z$  involved to fill a row, throughout the filling, processed over approximately 500 frames. Interestingly, it shows a bimodal distribution with a significant dispersion. One may speculate that the maximum of the probability distribution, located at $V_z^{max}=62\,mm/s$, corresponds to the wetting dynamics for an ideal sample. The large dispersion can be attributed to surface inhomogeneities resulting in a surface energy lowering (caused by defects or impurities). Larger defects might be the cause of the second peak, located around $V_z=105\,mm/s$. 

\begin{figure*}[t!]
\center
\includegraphics[width=15cm,keepaspectratio]{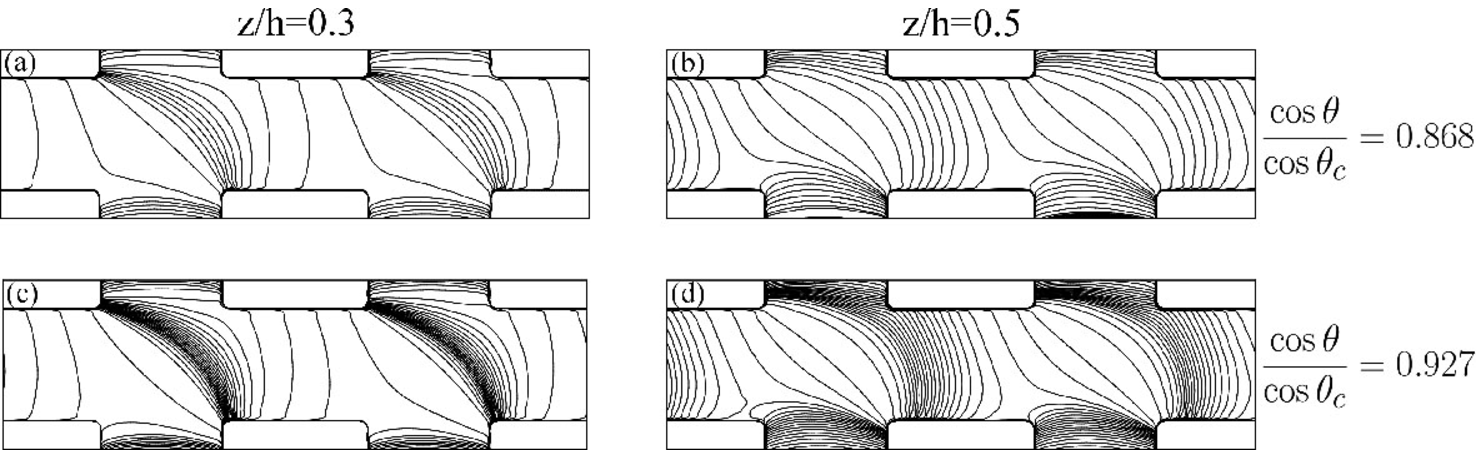}
\caption{\label{num} Numerical simulations reveal the details of the zipping mechanism. Two values of the wetting properties are used with respect to the critical angle given in equation (\ref{thetac}): $\cos \theta /\cos \theta_{c}=0.868$ ((a) and (b)) and $\cos \theta /\cos \theta_{c}=0.927$ ((c) and (d)). The interface position is tracked as a function of time (from left to right) at a given height $z$ ($z/h=0.3$ for (a) and (c) and $z/h=0.3$ for (b) and (d)) and all plots are equispaced with the same numerical time lag. When approaching the critical point ((c) and (d), (AVI movies, 0.2 Mb)) the zipping is experiencing a dynamical slowing down followed by a rapid merging process. The pulling mechanism provided by the liquid on the top of the pillars is inducing a complex $3d$ stretching that deforms the interface. An insight of this is provided by the cuts at different heights ($z/h=0.5$ and $z/h=0.3$). In particular, quantitative agreement is found with the experiments, assuming a height less than $h/2$ ($z/h=0.3$ in (c)).}
\end{figure*}

Beyond the observed dispersion of the zipping velocity, it is worthwhile noticing that rows do not fill one by one continuously. A schematic picture of this is provided in the inset at fig.~\ref{zipping10}(c). When the filled square is still small, a sideway row gets filled (at time $T_1$), and then the next row (at time $T_2$) after a latency time ($l_t=T_2-T_1>0$). The latency is decreasing with increasing square size until it finally vanishes (at about $270\,\mu m$ square side length, \ding{172} in fig.~\ref{zipping10}(c)). Afterwards, a different scenario is observed. The new row is already started (at time $T'_1$) before the former one has been completed (at time $T'_2$). In this case, we express the latency time as $l_t=T'_1-T'_2<0$ (\ding{173} in fig.~\ref{zipping10}(c)). As a consequence, the filled area cannot be seen as a perfect parallelepiped any longer although the macroscopic square-like shape remains, considering that the filled surface gets larger (see fig.~\ref{fig1}(d-f)). Moreover, the side length corresponding to $l_t=0$ can simply be calculated from velocities by $V_z^{max}/V_f\times d=270\,\mu m$, which is in good agreement with the length extrapolated in fig.~\ref{zipping10}(c).

Details of the front advancing in between the posts are shown in fig.~\ref{zipping100}(a). The height is {\it a priori} not exactly known. The dynamics of the interface are captured in a plane close to the middle of the pillars. Based on the objective specifications, we estimate it to be at $z=4\pm1\,\mu m$ from the base. The dynamics are processed from an experimental record chosen especially for its low $V_z/V_f$ ratio to better reveal main and zipping front progressions at once. Thus, when averaged over $20\,\mu m$, the main front velocity is $V_f=1.9\,mm/s$ and the zipping front velocity is $V_z=10\,mm/s$. The figure enables a comprehensive picture of the complex dynamics in the $2d$ plane: position, curvature and velocity variation of the interface are easily accessible. Front and zipping interface positions as a function of time are quantitatively tracked in fig.~\ref{zipping100}(b) along the straight lines passing through the middle of the 4 rows shown in fig.~\ref{zipping100}(a). First, the two perpendicular liquid-gas interfaces defined as main and zipping fronts spread between the pillars and reach almost simultaneously the edge. Next, they fastly merge, and the resulting interface straighten. Rectangles drawn in fig.~\ref{zipping100}(b) highlight this stage. In rectangles labelled ``1'', main and zipping fronts are superimposed, indicating that the interface at an intersection is almost symmetric. In rectangles labelled ``2'', the merged front has reached a lower velocity and an asymmetry develops: $V_z$ increases more than $V_f$. Afterwards, when the interface has reached the opposite post, it splits into 2 fronts that progress in main and zipping direction respectively, with drastically different velocities. The passage is made within $4\,ms$ for the main front whereas it is almost 2 orders of magnitude less for the zipping front.

Here the interface is described in one plane, but since the filling is a $3d$ process, it is straightforward that several planes should be analyzed simultaneously. Unfortunately this cannot be achieved experimentally. Therefore, we have carried out numerical simulations to better reveal the $3d$ structure as well as the dynamics of the zipping mechanism.

For the numerical simulations presented here, we use a $3d$ lattice Boltzmann \cite{LBM} model for single component multiphase flows whose details are discussed in \cite{Mauro06}. In all these mesoscale models \cite{Anderson98}, it is impossible to match all the physical parameters and preserve at the same time a feasible computation. Typically the width of the interface is too large and the liquid-gas density difference is unphysically small with respect to physical reality. As already noticed elsewhere \cite{Yeomans}, this may affect the speed of the interface that, if directly proportional to the interface width and inversely proportional to the density difference, will move too quickly in the simulations when compared to an experiment. This must be taken into account by renormalizing the time scale if one wants to reconstruct a correct pathway as a function of time \cite{Yeomans}.

All the geometrical aspect ratios for the surface micro-structures can be adapted in order to match those typically studied in the experiments \cite{Dupuis05} and wall interactions with wetting properties are introduced as explained in \cite{Mauro06}. The initial condition consists of a liquid reservoir on the top of the pillars and an infiltration characterized by a flat front perturbed with an advanced liquid precursor, just as experimentally observed at the onset of zipping. In this way the liquid precursor induces a lateral motion that is continuously driven by the pulling mechanism of surface tension on the top of the pillars. After a small time transient, the properties of this numerical zipping motion show a remarkable independence on the initial condition details and are reported in fig.~\ref{num} (zipping is advancing from left to right in these figures). Two wetting properties are shown, $\cos \theta /\cos \theta_{c}=0.868$ ((a) and (b)) and $\cos \theta /\cos \theta_{c}=0.927$ ((c) and (d)) and the interface contour is tracked in a two dimensional plane for two different heights, $z/h=0.3$ ((a) and (c)) and $z/h=0.5$ ((b) and (d)). 

\begin{figure}[t!]
\center
\includegraphics[width=7cm,keepaspectratio]{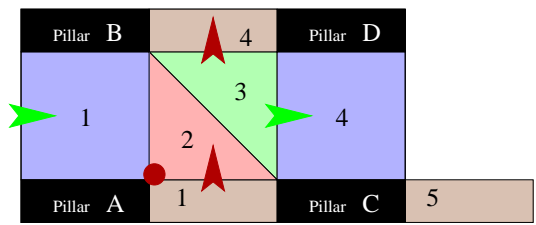}
\caption{\label{zip} Sketch of the different steps in the zipping process. The numbered areas refer to the text. As in figs.~\ref{zipping100}(a) and \ref{num}, the zipping front is from left to right (green arrows) and main front is from bottom to top (red arrows).}
\end{figure}

The results presented in figs.~\ref{zipping100}(a) and \ref{num} reveal the complex structure of the zipping process, both from experimental and numerical viewpoints. Firstly the liquid precursor fastly merges with its neighbor and after that experiences a passage through the walls. It is observed that as one approaches the critical point the numerical pathway at the height $z/h=0.3$ in fig.~\ref{num}(c) shares quantitative agreement with the experimental one. The main steps of the zipping process are sketched in fig.~\ref{zip} and should be superimposed to the above figures.

Starting with the case $z/h=0.3$, the successive steps are described as numbered in fig.~\ref{zip}.

1- The pulling mechanism induced at the top of the pillars makes zipping and main front interfaces merge (red point at the corner of pillar A).

2- The liquid bridge suspended on the four pillars collapses at pillar A, producing a local top-to-bottom pulling mechanism. The dynamics is dominated by surface tension effects that tend to reduce the strongly curved surface of the merged interface, by drawing liquid from the top. Wall effects are negligible in this extremely fast process (within $20\,\mu s$ in the experiment): it is almost independent of the wetting properties (compare for example case (b) and (d) in fig.~\ref{num}).

3- Since the interface is straight, the pulling persists but the interface above is behind (late). Viscous dissipation dominates, making the front much slower (time scale is $\approx 1\,ms$ in the experiment). Moreover, the interface is still pinned on the edges of pillars B and C.

4- The interface reaches pillar D and splits into two parts and then the filling progresses. The dynamical details are influenced by wall interactions that cause a slowing down for the propagation. This is strongly pronounced especially when the wetting properties approach the critical point passing from (a) to (c) in fig.~\ref{num}.

The pulling mechanism on the top of the pillars is stretching the interface, and obviously the closer to the top the stronger it is. Consequently, we choose a cut plan at $z/h=0.5$ to evidence the change in the dynamics. If we compare with $z/h=0.3$, the front is slower in region 2, but much faster in region 3 where the slowing down is not observed (compare (c) to (d) in fig.~\ref{num}). Interestingly, this makes the liquid invade the gap between pillars C and D at $z/h=0.5$ before than at $z/h=0.3$. The resulting stretching overcomes the wall interactions that, being in the slightly hydrophobic regime ($\theta_f \sim 100^{\circ}$), act to prevent a complete wetting of the boundaries (figs.~\ref{zipping100}a and \ref{num}c).

One has to remember that liquid is present in area 5, causing the interface that bridges over pillars C and D to be more curved than the one over pillars B and D (the portion beyond these pillars in not filled). This induces a faster filling in the zipping direction (green arrows in fig.~\ref{zip}) than in the main front direction (red arrows). The front velocity between pillars C and D is progressively increased while the interface along pillar C is deformed by a stronger pulling (fig.~\ref{num}(d)). It can be assumed that this effect is mainly at the base of the macroscopic square front and that the zipping velocity is quite sensitive to slight variations of the shape of the pillars top.

In summary, we have investigated experimentally and numerically details of the spreading dynamics for the CB to W transition on a micro-patterned substrate, close to the critical point above which filling occurs. The already wet portion expands by filling the rows along the pattern through a zipping mechanism, thus preserving at a macroscopic scale the structure of the underlying square lattice. We have shown that in this regime, even if the pattern grows in a regular way (the front velocity $V_f$ is almost contant), the zipping velocity $V_z$ involved to fill a row undergoes a noticeable dispersion (from $25$ to $125\,mm/s$ in a typical experiment). This reveals the strong sensitivity of the dynamics to local wetting properties without prejudging the cause (roughness variation, geometrical imperfection, material inhomogeneity, dust deposit, \ldots). Interestingly, this translates into a latency acting as a self-adjusting parameter and, consequently, makes the square shape robust at a macroscopic level. Interested in getting further details at a smaller scale, we have performed, to our knowledge, unprecedented high-speed submicron observations of the zipping-wetting dynamics. The dynamical details of the interface demonstrate the complexity of the invasion process, while being in very good agreement with numerical predictions for $z/h=0.3$ (corresponding to $z=3\,\mu m$ in the experiment). Moreover, the numerical results shown for $z/h=0.5$ give a deeper insight of the complex $3d$ stretching experienced by the interface. Finally, the analysis of the submicron dynamics allows us to provide a simplified explaination for the resulting wetting dynamics.

\acknowledgments

We gratefully acknowledges discussions with J.\ Yeomans. M.S.\ and B.M.B.\ thank STW (Nanoned Programme) and A.M.P.\ thanks Microned for financial support.


\end{document}